\documentclass[%
reprint,
superscriptaddress,
amsmath,amssymb,
aps,
]{revtex4-2}

\usepackage{graphicx}% Include figure files
\usepackage{dcolumn}% Align table columns on decimal point
\usepackage{bm}% bold math
\usepackage{hyperref}% add hypertext capabilities
\usepackage{pgf}
\usepackage{soul}  % to highlight some changes

\usepackage{times} %change font to times new roman
\usepackage{siunitx} % for SI units
\usepackage{physics} %%% braket notation  etc.
\usepackage{float} %Place the figure exactly here (requires \usepackage{float}).

\usepackage{pdfpages}

\makeatletter
\AtBeginDocument{\let\LS@rot\@undefined}
\makeatother

\begin{document}

\title{Bringing Dark States to Light after Frustrated Tunneling Ionization}

\author{Yu He}
\email{yuhe@mpi-hd.mpg.de}
\affiliation{Max-Planck-Institut für Kernphysik, Saupfercheckweg 1, 69117 Heidelberg, Germany}
\affiliation{Max-Born-Institut, Max-Born-Strasse 2A, 12489 Berlin, Germany}

\author{Eleni Tzima}
\affiliation{Max-Planck-Institut für Kernphysik, Saupfercheckweg 1, 69117 Heidelberg, Germany}

\author{Gergana D. Borisova}
\affiliation{Max-Planck-Institut für Kernphysik, Saupfercheckweg 1, 69117 Heidelberg, Germany}

\author{Noah L. Wach}
\affiliation{Max-Planck-Institut für Kernphysik, Saupfercheckweg 1, 69117 Heidelberg, Germany}

\author{Aaron T. Bondy}
\affiliation{Department of Physics and Astronomy, Drake University, Des Moines, Iowa 50311, USA}

\author{Juan C. Del Valle}
\affiliation{Instituto de Física, Universidad Nacional Autónoma de México, Ciudad de México 04510, Mexico}

\author{\\Klaus Bartschat}
\email{klaus.bartschat@drake.edu}
\affiliation{Department of Physics and Astronomy, Drake University, Des Moines, Iowa 50311, USA}

\author{Christian Ott}
\email{christian.ott@mpi-hd.mpg.de}
\affiliation{Max-Planck-Institut für Kernphysik, Saupfercheckweg 1, 69117 Heidelberg, Germany}
\author{Thomas Pfeifer}
\email{thomas.pfeifer@mpi-hd.mpg.de}
\affiliation{Max-Planck-Institut für Kernphysik, Saupfercheckweg 1, 69117 Heidelberg, Germany}

%\date{\today}

\begin{abstract}
High-harmonic generation (HHG) in gases is typically regarded as a sub-laser-cycle process, well described by the semiclassical three-step model. In this work, we report the extreme-ultraviolet (XUV) emission observed in an HHG experiment arising from a longer temporal process associated with dark states populated through frustrated tunneling ionization. The emission exhibits a larger spatial divergence than typical high-order harmonics, resembling the polarization decay of optically bright states. Supported by the time-frequency analysis of the theoretical calculations, we show that the emission originates from the nonradiative capture of the tunneling electrons into excited dark states of the neutral atom, followed by Floquet-like laser-dressing effects induced by the extended tail of the driving pulse. Our results reveal a largely unexplored pathway for XUV generation through laser-dressed dark states, thereby enabling the enhancement of coherent XUV emission at targeted wavelengths with high repetition rates.

\end{abstract}

\maketitle 
High-harmonic generation (HHG) serves as a cornerstone of strong-field physics and attosecond science. Microscopically, the \mbox{well-established} three-step model \cite{kulander1993dynamics, PhysRevLett.70.1599, PhysRevLett.71.1994} successfully describes the main characteristics of HHG in gases well above the ionization threshold, where the influence of the ionic potential can be neglected \cite{PhysRevA.49.2117}. The physical mechanism becomes more complex near and below the ionization threshold, and in the presence of resonances. The influence of bound Rydberg \cite{chini2014coherent, PhysRevA.92.013404, PhysRevLett.117.203001, PhysRevA.96.023418, yun2018coherent, Camp_2018, PhysRevLett.129.173202, zhao2024coherent, PhysRevLett.134.123203, pp22-qc27} and autoionization states \cite{Ganeev:06, Suzuki:06, PhysRevLett.104.123901,PhysRevLett.112.233002, Singh:21,fareed2017high, PhysRevLett.130.073201} on high-frequency light emission under HHG conditions has been the subject of many experimental and theoretical studies, revealing a wealth of phenomena beyond the description of the three-step picture. The resulting spectral features encode information on the dynamics of the electronic wave packet and offer opportunities for ultrafast spectroscopy \cite{doi:10.1126/science.1123904, smirnova2009high, shafir2009atomic}. In addition, they provide attractive prospects as coherent table-top light sources for applications requiring narrowband extreme-ultraviolet (XUV) radiation with high repetition rates \cite{yost2009vacuum}. 

One strong-field process closely related to HHG, which underlies the formation of bound excited states, is frustrated tunneling ionization (FTI) \cite{PhysRevLett.101.233001}. FTI describes the phenomenon that a substantial fraction of neutral atoms survives in the excited states after being exposed to external laser fields well in the strong-field tunneling regime. The discovery has prompted a series of experimental \cite{PhysRevLett.109.093001, PhysRevLett.110.203002, PhysRevLett.118.013003} and theoretical investigations \cite{Landsman_2013, Zhao:19, Liu:21, PhysRevA.107.043106} owing to its fundamental importance in strong-field physics. Moreover, it has also enabled novel applications, including the acceleration of neutral atoms \cite{eichmann2009acceleration} as well as the generation of coherent XUV emission \cite{yun2018coherent, PhysRevLett.134.123203}.

As a tunneling-plus-rescattering process, FTI can populate excited states of both the same and opposite parity relative to the ground state. Dark states refer to those states that are optically forbidden from the ground state by single-photon transitions. Typically long-lived and thus metastable, they have widespread impact in physics, including the creation of population inversion for the development of lasers \cite{maiman1960stimulated, PhysRevLett.6.106}, optical clocks \cite{RevModPhys.87.637}, electromagnetically induced transparency \cite{PhysRevLett.64.1107, PhysRevLett.66.2593, RevModPhys.77.633}, and ultrafast laser coupling among quantum states \cite{ott2014reconstruction, PhysRevLett.112.103001, PhysRevLett.127.023202, Leshchenko:23, he2025bringing}. However, since the radiative transitions between the ground and dark states are forbidden due to the dipole selection rule, such nonradiative states are generally assumed to be inactive in HHG experiments \cite{PhysRevLett.130.073201}. Recently, Zhao {\it et al.} reported the experimental observation of an XUV supercontinuum spatially separated from the HHG spectrum, which was interpreted by a joint ac Stark and dark-to-bright emission mechanism \cite{zhao2024coherent}.

In this Letter, we investigate the largely unexplored role of dark states in XUV emission under HHG conditions. We report the experimental identification of an XUV-emission channel from virtual laser-dressed dark states following FTI, in analogy to Floquet states \cite{CHU20041, Wu_2016}. Theoretical analysis provides a time-dependent perspective on the underlying dynamics, and confirms the long tail of the driving pulse after the main envelope as the key ingredient enabling this otherwise dipole-forbidden emission channel. These findings reveal the active role of dark states in strong-field light-matter interactions and underscore the importance of the extended laser tail, which is commonly present in few-cycle laser pulses. Our work opens avenues for the ultrafast preparation and control of \mbox{Rydberg} wave packets, as well as the enhancement of XUV emission at targeted wavelengths by tailoring the intensity and spectral composition of the tail pulse.

\begin{figure}
\includegraphics[width=0.45\textwidth]{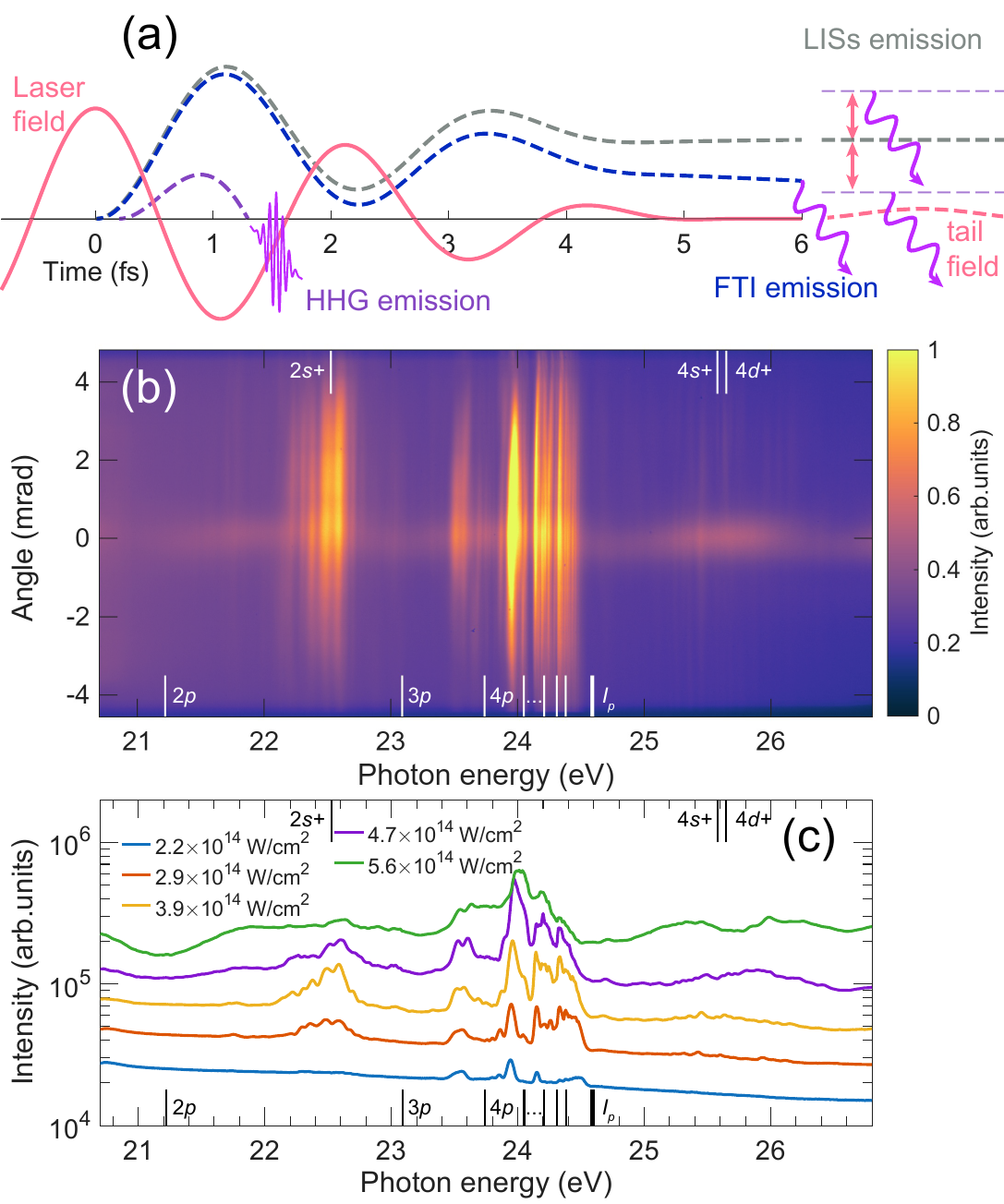}
\caption{\label{fig1} \textbf{Scheme and experimental results.} (a) Illustration of the electron trajectories for three sets of XUV emission: HHG (purple curve), FTI emission from the polarization decay of bright states (blue curve), and LISs from dark states (gray curve). Dark excited states can be populated through FTI. By absorption or emission of a VIS photon in the presence of the residual tail field, the Floquet-like LISs become dipole-coupled to the ground state and can radiate light. (b) Spatially and spectrally resolved XUV spectrum generated in helium at a peak driving-pulse intensity of $3.9\times10^{14}$~$\si{W/cm^{2}}$. In contrast to the weak HHG emission in a small spatial divergence range, FTI and LIS emission features are more pronounced and divergent. (c) Spatially integrated XUV intensity for different driving-pulse intensities. The field-free photon energies of the $1snp$ bright states and the expected photon energies of the laser-dressed dark states are marked by vertical lines in (b) and (c).}
\end{figure}

Figure~\ref{fig1}(a) illustrates the schematic process considered in this work. In the three-step model, an electron wave packet is initiated by tunnel ionization when neutral atoms are exposed to a strong laser field. Based on classical simulations of the free-electron trajectories, electrons that tunnel out after the peak of the laser field can recombine with the parent ion and give rise to HHG emission. By contrast, electrons liberated near the laser peak are less likely to recombine with the parent ion, but instead can be nonradiatively captured into excited states through FTI. This process facilitates the population of both bright and \textit{dark} excited states. Narrowband coherent XUV radiation can be emitted through polarization decay as a result of the coherence between the ground and bright states. Driving HHG and FTI in helium with an optical pulse typically requires peak pulse intensities well above $1\times10^{14}$~$\si{W/cm^{2}}$ (for reference, the Keldysh parameter \cite{keldysh1965ionization} $\gamma \approx 1$ at $3\times10^{14}$~$\si{W/cm^{2}}$ for a laser wavelength of 650 nm). However, realistic experimental laser pulses inevitably feature a long tail with an intensity of a few percent of that of the main pulse. Based on the fact that light-induced states (LISs) have been reported at intensities on the order of $10^{12}$~$\si{W/cm^{2}}$ in transient absorption studies in helium \cite{PhysRevA.86.063408, chini2013sub, Lucchini_2013, PhysRevA.92.033408, kfjh-zc96, Yue:26}, it is expected that the signatures of XUV emission from laser-dressed dark states following FTI will appear in HHG experiments. 

In the present experiment, a commercial laser system (FEMTOPOWER$^{\rm{TM}}$ HE/HR CEP4) delivers 20 fs, 2.5~mJ pulses with a central wavelength of 780 nm at a repetition rate of 3 kHz. Few-cycle visible (VIS) laser pulses (5 fs duration, $\sim$1~mJ pulse energy) are generated through spectral broadening in a hollow-core fiber followed by pulse compression via chirped mirrors. The VIS spectrum is shown in the Supplemental Material \cite{SM_He}; its central wavelength is around 650 nm. The VIS pulses are focused into a 3-mm-long gas cell filled with helium with a backing pressure of $\sim$500 mbar to drive HHG. More details on the experimental setup are described in Refs.~\cite{10.1063/1.5091069, He:26}. A motorized iris is used to control the driving-pulse intensity. After removing the fundamental beam by an aluminum filter, the generated XUV beam is dispersed by a grating (Hitachi, 001-0640) and detected by a charge-coupled device (CCD) camera (Princeton Instruments, PIXIS-XO: 400B). 

Figure~\ref{fig1}(b) shows the recorded two-dimensional XUV image in the vicinity of singly excited states of helium, at a peak driving-pulse intensity of $3.9\times10^{14}$~$\si{W/cm^{2}}$. On top of the weak HHG emission that is limited to a small divergence range of $\pm$0.5~mrad, narrowband XUV emission shows up over a large divergence range. The emission near the $1snp$ singly excited states has previously been observed and identified as polarization decay of bright states following FTI \cite{yun2018coherent}. While the red shift of the FTI emission with respect to the field-free state energies is a macroscopic effect (see the measured XUV spectra for different target pressures in the Supplemental Material \cite{SM_He}) that can be attributed to phase matching \cite{chini2014coherent}, the spectral sidebands away from any bright states around 22.5 and 25.6 eV can only be explained by LIS emission. The LIS and FTI emissions display a similarly large divergence, which is a characteristic feature of the FTI process \cite{yun2018coherent}. As FTI electrons travel longer trajectories than those in HHG, they accumulate a larger phase and are thus more sensitive to the variation of the driving laser pulse \cite{PhysRevLett.100.143902}. Since the nonresonant XUV emission is significantly weaker than the FTI and LIS features, the contribution of excited states populated via the absorption of one XUV HHG photon is inconsiderable. 

Figure~\ref{fig1}(c) presents the spatially integrated XUV spectra measured for different driving-pulse intensities. At the lowest intensity, only FTI emission from the bright Rydberg states is visible (red-shifted due to macroscopic propagation, see Supplemental Material \cite{SM_He}). These emission features are enhanced with increasing driving-pulse intensities, along with the appearance of sideband structures, which we attribute to LIS emission. Referring to the transient virtual states formed by absorbing or emitting an additional VIS photon from a real state $1snl$ as $1snl\pm$, the sideband structure around 22.5 eV can be predominantly designated as $1s2s+$; the weaker sideband above the ionization potential around 25.6 eV is mainly attributed to $1s4s+$ and $1s4d+$. We note that the recorded XUV spectra also contain contributions from other LIS emission channels, including $1s4s-$ and $1s4d-$ around 21.8 eV, $1s6s-$ and $1s6d-$ around 22.3 eV, and $1s6s+$ and $1s6d+$ around 26.1 eV, respectively. These XUV sidebands are analogous to the observation of laser-induced two-photon and anti-Stokes scattering initiated by a glow discharge \cite{10.1063/1.89756, PhysRevLett.40.1493} or electron collisions \cite{CHENG1993183}.

\begin{figure}
\includegraphics[width=0.45\textwidth]{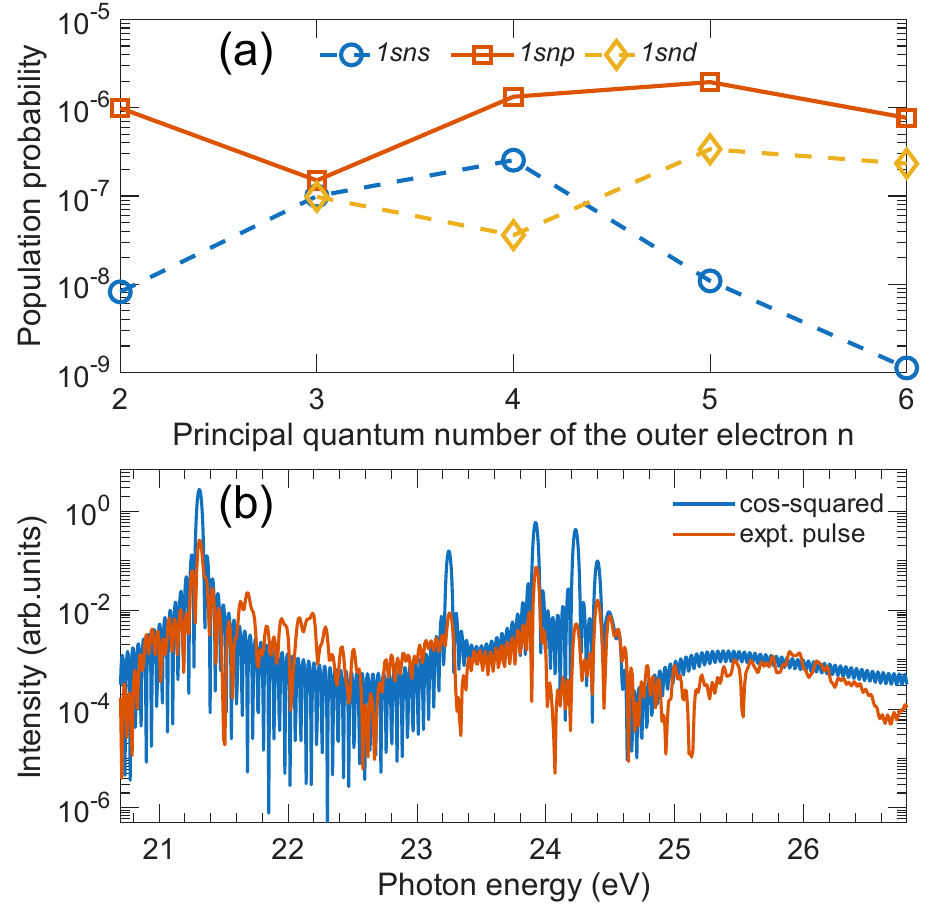}
\caption{\label{fig2} \textbf{Numerical TDSE calculation results.} (a) Population probability of the excited states after interaction with the cos-squared driving pulse. A significant amount of $1sns$ and $1snd$ states are also populated, albeit their transition from the ground state $1s^2$ is dipole-forbidden. (b) XUV spectra in the vicinity of the singly excited states of helium calculated for cos-squared and experimental pulses. While the clean cos-squared pulse only leads to sharp FTI emission peaks associated with the bright states, the realistic experimental pulse results in additional LIS features in between $1s2p$ and $1s3p$ states, and above the ionization threshold.}
\end{figure}

\begin{figure*}[!htbp]
\includegraphics[width=0.9\textwidth]{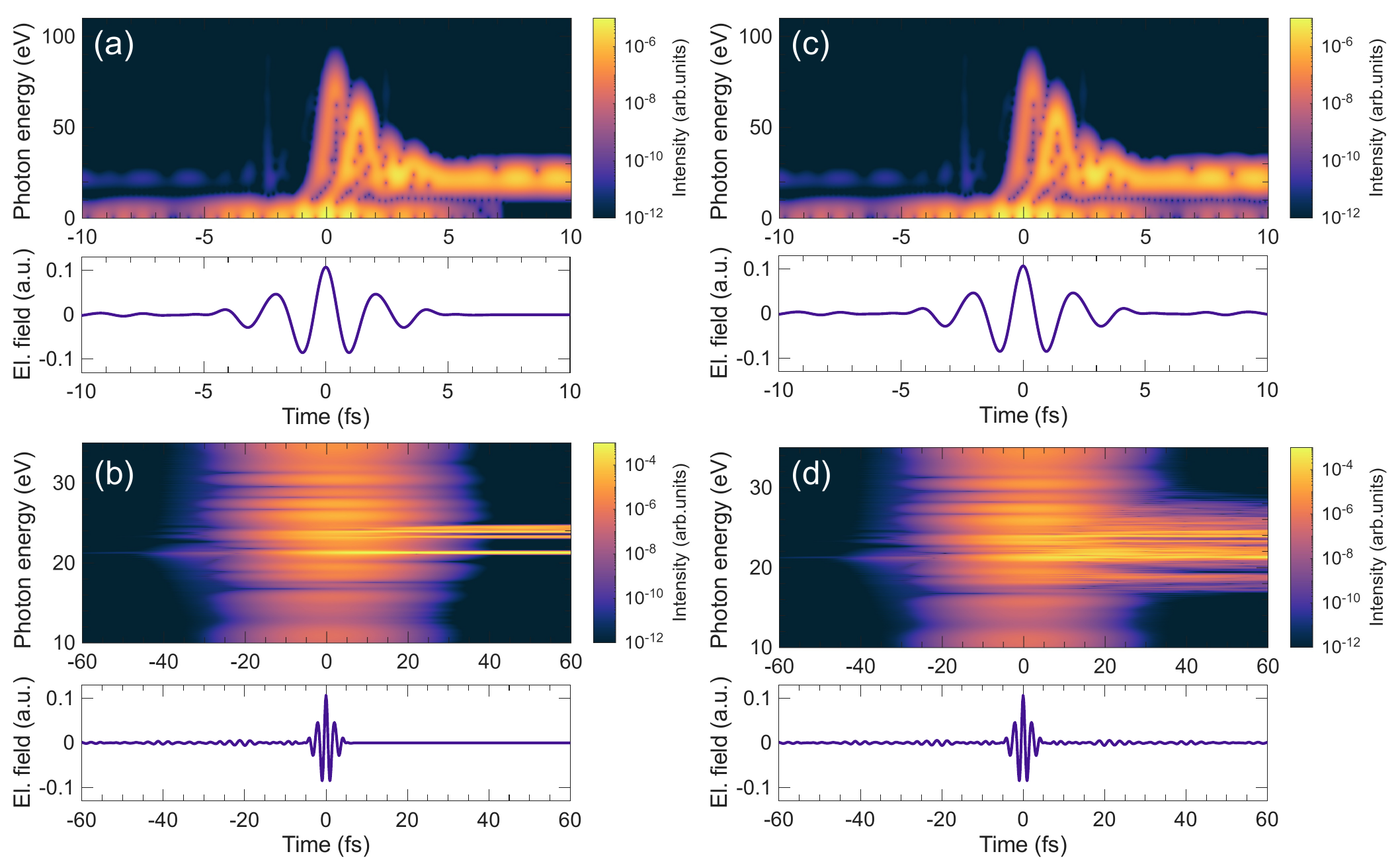}
\caption{\label{fig3} \textbf{Time-frequency analysis of the dipole emission.} (a, c) Gabor spectrogram of the dipole acceleration using a 0.2-optical-cycle (434-as) FWHM Gaussian window function. Attosecond structure of the branches from the short and long trajectories in HHG emission is resolved. Persistent emission features appear after the main laser envelope, but are not spectrally well resolved. (b, d) Gabor analysis using a 10-optical-cycle (21.7-fs) FWHM Gaussian window function. FTI emission from bright states is present in both panels, while the additional structures due to laser-dressed dark states only appear in (d). The lower panels show the laser electric field in atomic units (a.u.) used in the calculations [realistic experimental pulse in (c) and (d), the same pulse without the post-pulse (smoothly truncated at 6 fs over a 1-fs window) in (a) and (b)].}
\end{figure*}

To support the interpretation of the experimental results and pinpoint the role of the laser tail, we perform numerical calculations based on solving the time-dependent Schrödinger equation (TDSE) within the single-active-electron (SAE) approximation \cite{Birk_2020, PhysRevA.102.062809}. Details regarding the potential used, the propagation scheme, numerical parameters, and convergence checks performed can be found in the second part of the Appendix in Ref.~\cite{Birk_2020}.  Since the earlier works, the numerical efficiency was improved by using a variable radial grid as described in Ref.~\cite{atoms12070034}. This variable grid greatly simplifies calculations with long propagation times. The SAE-TDSE calculation aims to capture the microscopic single-atom dynamics, while macroscopic effects such as phase matching and propagation effects are not considered. The modeled helium atom is subjected to a laser pulse with a cos-squared envelope and an intermediate peak intensity of $4\times10^{14}$~$\si{W/cm^{2}}$. We choose a central wavelength of 650 nm and a full-width-at-half-maximum (FWHM) duration of 4 fs to approximate the overall bandwidth of the experimental spectrum. The simulated populations of the excited states after the end of the driving pulse are shown in Fig.~\ref{fig2}(a). In addition to the bright $1snp$ states that have been revealed by previous studies, there are notable populations in the dark $1sns$ and $1snd$ states. In the simulated HHG spectrum shown in Fig.~\ref{fig2}(b) (blue line), sharp resonant peaks corresponding to the polarization decay of the bright $1snp$ states are presented.

To mimic the experimental pulse and qualitatively reproduce the observed LIS emission sidebands, a more realistic laser pulse constructed by taking the measured driving laser spectrum and assuming a flat spectral phase with a peak intensity of $4\times10^{14}$~$\si{W/cm^{2}}$ is also employed (see Supplemental Material \cite{SM_He} for more details). When the target system is driven by this laser pulse with a long temporal tail [shown in the lower panels of Figs.~\ref{fig3}(c) and \ref{fig3}(d)], additional emission features in between the bright states and above the ionization threshold are observed in the calculated XUV spectrum shown in Fig.~\ref{fig2}(b) (red line). The emergence of these structures in the presence of a long laser tail confirms the LIS interpretation. 

For a time-dependent perspective on the roles of excited states in the HHG process, we perform a time-frequency Gabor analysis \cite{PhysRevA.51.R1750} of the dipole acceleration obtained from the TDSE simulations. We first consider a driving pulse identical to the constructed experimental pulse but with the post-pulse removed by truncating the field at 6 fs (over a 1-fs smooth window), as shown in the lower panels of Figs.~\ref{fig3}(a) and \ref{fig3}(b). By employing a short Gaussian window with a FWHM duration of 0.2 optical cycles (434 as), we resolve the attosecond temporal structure of the produced XUV emission. As presented in Fig.~\ref{fig3}(a), the spectrogram shows the branches associated with the short and long trajectories in HHG emission \cite{PhysRevA.49.2117, PhysRevLett.117.203001}. These branches disappear as the field envelope approaches its end at $7$ fs, after which persistent emission features emerge. However, the limited energy resolution prevents the narrowband resonant structures from being resolved. To resolve these structures, the Gaussian window in the Gabor analysis is increased to 10 optical cycles (21.7 fs), and the FTI emission manifests itself as sharp spectral structures as shown in Fig.~\ref{fig3}(b).

For the calculations using the constructed experimental pulse shown in the lower panels of Figs.~\ref{fig3}(c) and \ref{fig3}(d), the spectrogram obtained with a 434-as FWHM Gaussian window in Fig.~\ref{fig3}(c) shows HHG branches within the main field envelope and persistent emission features afterwards, closely resembling those in Fig.~\ref{fig3}(a). In contrast, the spectrogram obtained with a 21.7-fs FWHM Gaussian window in Fig.~\ref{fig3}(d) displays more resonant structures as compared to Fig.~\ref{fig3}(b). The additional emission peaks are attributed to LIS features. The comparison of the two sets of calculations clearly demonstrates the importance of a long tail in the experimental pulse in the observation of the LIS sidebands, which are only active in the presence of a non-zero dressing electric field. In addition, these results indicate that recapturing of electrons into the excited states in FTI occurs predominantly near the end of the main pulse envelope, consistent with the experimental study reported recently \cite{PhysRevLett.134.123203}.

\begin{figure}
\includegraphics[width=0.45\textwidth]{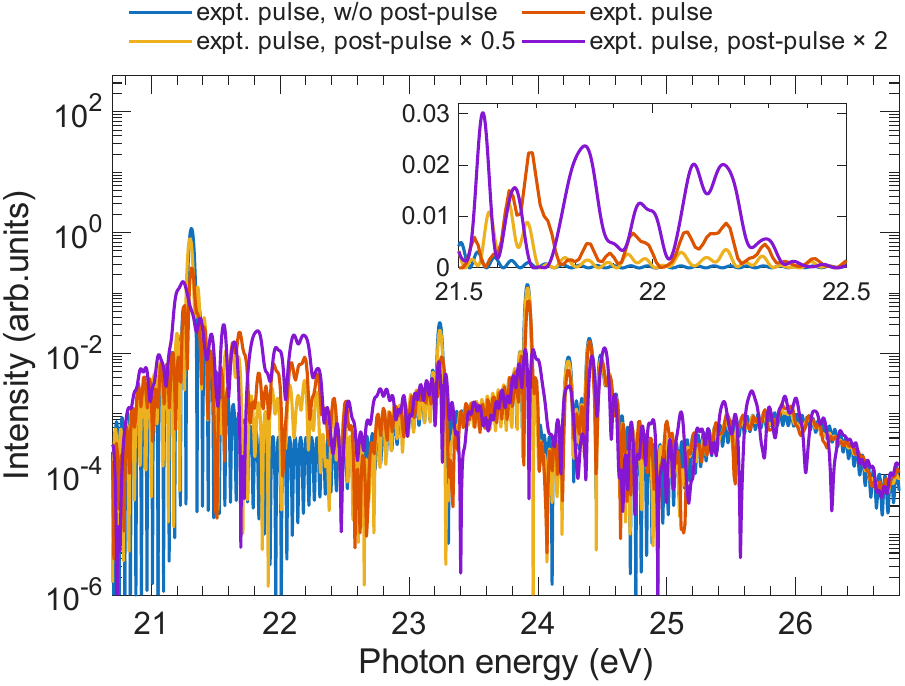}
\caption{\label{fig4} \textbf{Influence of the post-pulse on XUV emission.} Calculated XUV spectra for four different driving fields: the experimental pulse [as in Figs.~\ref{fig3}(c) and \ref{fig3}(d)], the experimental pulse with the post-pulse removed [as in Figs.~\ref{fig3}(a) and \ref{fig3}(b)], and the experimental pulse with the post-pulse electric-field amplitude scaled by factors of 0.5 and 2. The LIS emission, particularly in the vicinity of 22 eV, becomes more pronounced with increasing post-pulse amplitude, as shown in the inset plotted on a linear scale.}
\end{figure}

Finally, we investigate the influence of the post-pulse on XUV emission by varying its amplitude. Following the preparation of the Rydberg wave packets through FTI by the main pulse, the post-pulse can change the populations of the excited states via ionization and resonant transitions \cite{yun2018coherent}. As shown in Fig.~\ref{fig4}, compared to the spectrum in which the post-pulse of the driving pulse is removed, the presence of the post-pulse and its increasing amplitude leads to substantial modification of the FTI emission from bright states and LIS emission. In particular, the LIS emission around 22 eV is notably enhanced with increasing post-pulse amplitude, suggesting a route to controlling this emission channel by tailoring the residual field.

In summary, we have observed XUV emission sidebands resulting from laser-dressed dark states after FTI in helium. These results shed light on the widely unexplored role of dipole-forbidden states in HHG experiments, which are brought to light by virtue of the residual tail of the driving field. In contrast to the FTI emission from bright states that persists over the dephasing time governed by phase matching in the macroscopic medium, LIS emission is active only in the presence of a non-zero electric field of the dressing tail of the laser pulse. By tailoring this field, one can expect to control the LIS emission both temporally and spectrally (see Fig.~S3 in the Supplemental Material \cite{SM_He} for the calculated XUV spectra for different post-pulse wavelengths). These findings are relevant to experiments using compressed Yb-based laser sources, which have been gaining increasing importance in ultrafast and strong-field physics \cite{10.1063/5.0255819, chien2026filamentation, gao2026bright}, but whose nonlinear compression from a relatively long duration (over 100 fs) down to few- to single-cycle regime can leave substantial residual pedestals \cite{doi:10.1126/sciadv.abo1945, Pi:25}.

\textbf{Acknowledgment.} We thank Drs. Hao Liang, Xiao-Min Tong, and Mette Gaarde for fruitful discussions. The work of ATB, JCDV, and KB was supported by the National Science Foundation under grant No. PHY-2408484 and by computational resources through the ACCESS allocation No. PHY-090031.

\bibliography{helium}% Produces the bibliography via BibTeX.

@PREAMBLE{
 "\providecommand{\noopsort}[1]{}" 
 # "\providecommand{\singleletter}[1]{#1}%" 
}

@article{keldysh1965ionization,
  title={Ionization in the field of a strong electromagnetic wave},
  author={Keldysh, L V},
  journal={Sov. Phys. JETP},
  volume={20},
  number={5},
  pages={1307--1314},
  year={1965}
}

@article{Wu_2016,
	doi = {10.1088/0953-4075/49/6/062003},
	url = {https://doi.org/10.1088%2F0953-4075%2F49%2F6%2F062003},
	year = 2016,
	month = {feb},
	publisher = {{IOP} Publishing},
	volume = {49},
	number = {6},
	pages = {062003},
	author = {Mengxi Wu and Shaohao Chen and Seth Camp and Kenneth J Schafer and Mette B Gaarde},
	title = {Theory of strong-field attosecond transient absorption},
	journal = {Journal of Physics B: Atomic, Molecular and Optical Physics}
}

@article{PhysRevA.86.063408,
  title = {Light-induced states in attosecond transient absorption spectra of laser-dressed helium},
  author = {Chen, Shaohao and Bell, M. Justine and Beck, Annelise R. and Mashiko, Hiroki and Wu, Mengxi and Pfeiffer, Adrian N. and Gaarde, Mette B. and Neumark, Daniel M. and Leone, Stephen R. and Schafer, Kenneth J.},
  journal = {Phys. Rev. A},
  volume = {86},
  issue = {6},
  pages = {063408},
  numpages = {5},
  year = {2012},
  month = {Dec},
  publisher = {American Physical Society},
  doi = {10.1103/PhysRevA.86.063408},
  url = {https://link.aps.org/doi/10.1103/PhysRevA.86.063408}
}

@article{chini2013sub,
  title={Sub-cycle oscillations in virtual states brought to light},
  author={Chini, Michael and Wang, Xiaowei and Cheng, Yan and Wu, Yi and Zhao, Di and Telnov, Dmitry A and Chu, Shih-I and Chang, Zenghu},
  journal={Scientific reports},
  volume={3},
  number={1},
  pages={1105},
  year={2013},
url = {https://www.nature.com/articles/srep01105},
  publisher={Nature Publishing Group UK London}
}

@article{PhysRevA.92.033408,
  title = {Polarization control of absorption of virtual dressed states in helium},
  author = {Reduzzi, Maurizio and Hummert, Johan and Dubrouil, Antoine and Calegari, Francesca and Nisoli, Mauro and Frassetto, Fabio and Poletto, Luca and Chen, Shaohao and Wu, Mengxi and Gaarde, Mette B. and Schafer, Kenneth and Sansone, Giuseppe},
  journal = {Phys. Rev. A},
  volume = {92},
  issue = {3},
  pages = {033408},
  numpages = {8},
  year = {2015},
  month = {Sep},
  publisher = {American Physical Society},
  doi = {10.1103/PhysRevA.92.033408},
  url = {https://link.aps.org/doi/10.1103/PhysRevA.92.033408}
}

@article{PhysRevLett.112.103001,
  title = {Extracting Phase and Amplitude Modifications of Laser-Coupled Fano Resonances},
  author = {Kaldun, Andreas and Ott, Christian and Bl\"attermann, Alexander and Laux, Martin and Meyer, Kristina and Ding, Thomas and Fischer, Andreas and Pfeifer, Thomas},
  journal = {Phys. Rev. Lett.},
  volume = {112},
  issue = {10},
  pages = {103001},
  numpages = {5},
  year = {2014},
  month = {Mar},
  publisher = {American Physical Society},
  doi = {10.1103/PhysRevLett.112.103001},
  url = {https://link.aps.org/doi/10.1103/PhysRevLett.112.103001}
}

@article{PhysRevLett.127.023202,
  title = {Autoionizing Polaritons in Attosecond Atomic Ionization},
  author = {Harkema, N. and Cariker, C. and Lindroth, E. and Argenti, L. and Sandhu, A.},
  journal = {Phys. Rev. Lett.},
  volume = {127},
  issue = {2},
  pages = {023202},
  numpages = {5},
  year = {2021},
  month = {Jul},
  publisher = {American Physical Society},
  doi = {10.1103/PhysRevLett.127.023202},
  url = {https://link.aps.org/doi/10.1103/PhysRevLett.127.023202}
}

@article{Birk_2020,
doi = {10.1088/1361-6455/ab7c3f},
url = {https://dx.doi.org/10.1088/1361-6455/ab7c3f},
year = {2020},
month = {may},
publisher = {IOP Publishing},
volume = {53},
number = {12},
pages = {124002},
author = {Paul Birk and Veit Stooß and Maximilian Hartmann and Gergana D Borisova and Alexander Blättermann and Tobias Heldt and Klaus Bartschat and Christian Ott and Thomas Pfeifer},
title = {Attosecond transient absorption of a continuum threshold},
journal = {Journal of Physics B: Atomic, Molecular and Optical Physics}
}

@article{10.1063/1.5091069,
    author = {Stooß, Veit and Hartmann, Maximilian and Birk, Paul and Borisova, Gergana D. and Ding, Thomas and Blättermann, Alexander and Ott, Christian and Pfeifer, Thomas},
    title = "{XUV-beamline for attosecond transient absorption measurements featuring a broadband common beam-path time-delay unit and in situ reference spectrometer for high stability and sensitivity}",
    journal = {Review of Scientific Instruments},
    volume = {90},
    number = {5},
    pages = {053108},
    year = {2019},
    month = {05},
    issn = {0034-6748},
    doi = {10.1063/1.5091069},
    url = {https://doi.org/10.1063/1.5091069}
}

@article{ott2014reconstruction,
  title={Reconstruction and control of a time-dependent two-electron wave packet},
  author={Ott, Christian and Kaldun, Andreas and Argenti, Luca and Raith, Philipp and Meyer, Kristina and Laux, Martin and Zhang, Yizhu and Bl{\"a}ttermann, Alexander and Hagstotz, Steffen and Ding, Thomas and others},
  journal={Nature},
  volume={516},
  number={7531},
  pages={374--378},
  year={2014},
  url={https://www.nature.com/articles/nature14026},
  publisher={Nature Publishing Group UK London}
}

@article{PhysRevLett.70.1599,
  title = {Above threshold ionization beyond the high harmonic cutoff},
  author = {Schafer, K. J. and Yang, Baorui and DiMauro, L. F. and Kulander, K. C.},
  journal = {Phys. Rev. Lett.},
  volume = {70},
  issue = {11},
  pages = {1599--1602},
  numpages = {0},
  year = {1993},
  month = {Mar},
  publisher = {American Physical Society},
  doi = {10.1103/PhysRevLett.70.1599},
  url = {https://link.aps.org/doi/10.1103/PhysRevLett.70.1599}
}

@article{PhysRevLett.71.1994,
  title = {Plasma perspective on strong field multiphoton ionization},
  author = {Corkum, P. B.},
  journal = {Phys. Rev. Lett.},
  volume = {71},
  issue = {13},
  pages = {1994--1997},
  numpages = {0},
  year = {1993},
  month = {Sep},
  publisher = {American Physical Society},
  doi = {10.1103/PhysRevLett.71.1994},
  url = {https://link.aps.org/doi/10.1103/PhysRevLett.71.1994}
}

@article{PhysRevLett.112.233002,
  title = {Enhancing the Macroscopic Yield of Narrow-Band High-Order Harmonic Generation by Fano Resonances},
  author = {Rothhardt, Jan and H\"adrich, Steffen and Demmler, Stefan and Krebs, Manuel and Fritzsche, Stephan and Limpert, Jens and T\"unnermann, Andreas},
  journal = {Phys. Rev. Lett.},
  volume = {112},
  issue = {23},
  pages = {233002},
  numpages = {5},
  year = {2014},
  month = {Jun},
  publisher = {American Physical Society},
  doi = {10.1103/PhysRevLett.112.233002},
  url = {https://link.aps.org/doi/10.1103/PhysRevLett.112.233002}
}

@article{yun2018coherent,
  title={Coherent extreme-ultraviolet emission generated through frustrated tunnelling ionization},
  author={Yun, Hyeok and Mun, Je Hoi and Hwang, Sung In and Park, Seung Beom and Ivanov, Igor A and Nam, Chang Hee and Kim, Kyung Taec},
  journal={Nature Photonics},
  volume={12},
  number={10},
  pages={620--624},
  year={2018},
  url={https://www.nature.com/articles/s41566-018-0255-8},
  publisher={Nature Publishing Group UK London}
}

@article{PhysRevA.92.013404,
  title = {Interplay between resonant enhancement and quantum path dynamics in harmonic generation in helium},
  author = {Camp, Seth and Schafer, Kenneth J. and Gaarde, Mette B.},
  journal = {Phys. Rev. A},
  volume = {92},
  issue = {1},
  pages = {013404},
  numpages = {8},
  year = {2015},
  month = {Jul},
  publisher = {American Physical Society},
  doi = {10.1103/PhysRevA.92.013404},
  url = {https://link.aps.org/doi/10.1103/PhysRevA.92.013404}
}

@article{chini2014coherent,
  title={Coherent phase-matched VUV generation by field-controlled bound states},
  author={Chini, Michael and Wang, Xiaowei and Cheng, Yan and Wang, He and Wu, Yi and Cunningham, Eric and Li, Peng-Cheng and Heslar, John and Telnov, Dmitry A and Chu, Shih-I and others},
  journal={Nature Photonics},
  volume={8},
  number={6},
  pages={437--441},
  year={2014},
  url={https://www.nature.com/articles/nphoton.2014.83},
  publisher={Nature Publishing Group UK London}
}

@article{smirnova2009high,
  title={High harmonic interferometry of multi-electron dynamics in molecules},
  author={Smirnova, Olga and Mairesse, Yann and Patchkovskii, Serguei and Dudovich, Nirit and Villeneuve, David and Corkum, Paul and Ivanov, Misha Yu},
  journal={Nature},
  volume={460},
  number={7258},
  pages={972--977},
  year={2009},
  url={https://www.nature.com/articles/nature08253},
  publisher={Nature Publishing Group UK London}
}

@article{doi:10.1126/science.1123904,
author = {S. Baker  and J. S. Robinson  and C. A. Haworth  and H. Teng  and R. A. Smith  and C. C. Chirilă  and M. Lein  and J. W. G. Tisch  and J. P. Marangos },
title = {Probing Proton Dynamics in Molecules on an Attosecond Time Scale},
journal = {Science},
volume = {312},
number = {5772},
pages = {424-427},
year = {2006},
doi = {10.1126/science.1123904},
URL = {https://www.science.org/doi/abs/10.1126/science.1123904}}

@article{fareed2017high,
  title={High-order harmonic generation from the dressed autoionizing states},
  author={Fareed, MA and Strelkov, VV and Thir{\'e}, N and Mondal, Sudipta and Schmidt, BE and L{\'e}gar{\'e}, F and Ozaki, T},
  journal={Nature communications},
  volume={8},
  number={1},
  pages={16061},
  year={2017},
  url={https://www.nature.com/articles/ncomms16061},
  publisher={Nature Publishing Group UK London}
}

@article{Singh:21,
author = {Mangaljit Singh and Muhammad Ashiq Fareed and Vasily Strelkov and Alexei N. Grum-Grzhimailo and Alexander Magunov and Antoine Laram\'{e}e and Fran\c{c}ois L\'{e}gar\'{e} and Tsuneyuki Ozaki},
journal = {Optica},
number = {8},
pages = {1122--1125},
publisher = {Optica Publishing Group},
title = {Intense quasi-monochromatic resonant harmonic generation in the multiphoton ionization regime},
volume = {8},
month = {Aug},
year = {2021},
url = {https://opg.optica.org/optica/abstract.cfm?URI=optica-8-8-1122},
doi = {10.1364/OPTICA.434185},
}

@article{PhysRevLett.130.073201,
  title = {Ultrafast Resonant State Formation by the Coupling of Rydberg and Dark Autoionizing States},
  author = {Singh, Mangaljit and Fareed, Muhammad Ashiq and Birulia, Valeryia and Magunov, Alexander and Grum-Grzhimailo, Alexei N. and Lassonde, Philippe and Laram\'ee, Antoine and Marcelino, Romain and Shirinabadi, Ramin Ghahri and L\'egar\'e, Fran\ifmmode \mbox{\c{c}}\else \c{c}\fi{}ois and Ozaki, Tsuneyuki and Strelkov, Vasily},
  journal = {Phys. Rev. Lett.},
  volume = {130},
  issue = {7},
  pages = {073201},
  numpages = {6},
  year = {2023},
  month = {Feb},
  publisher = {American Physical Society},
  doi = {10.1103/PhysRevLett.130.073201},
  url = {https://link.aps.org/doi/10.1103/PhysRevLett.130.073201}
}

@article{PhysRevLett.104.123901,
  title = {Role of Autoionizing State in Resonant High-Order Harmonic Generation and Attosecond Pulse Production},
  author = {Strelkov, V.},
  journal = {Phys. Rev. Lett.},
  volume = {104},
  issue = {12},
  pages = {123901},
  numpages = {4},
  year = {2010},
  month = {Mar},
  publisher = {American Physical Society},
  doi = {10.1103/PhysRevLett.104.123901},
  url = {https://link.aps.org/doi/10.1103/PhysRevLett.104.123901}
}

@article{he2025bringing,
  title={Bringing weak transitions to light},
  author={He, Yu and Tong, Xiao-Min and Hu, Shuyuan and Borisova, Gergana D and Liang, Hao and Hartmann, Maximilian and Stoo{\ss}, Veit and Lyu, Chunhai and Harman, Zolt{\'a}n and Keitel, Christoph H and others},
  journal={Nature Communications},
  volume={16},
  number={1},
  pages={5322},
  year={2025},
  url={https://www.nature.com/articles/s41467-025-60701-9},
  publisher={Nature Publishing Group UK London}
}

@article{PhysRevLett.134.123203,
  title = {Resolving Recapture Dynamics of Rydberg Electrons via Laser-Driven Frustrated Tunneling Ionization},
  author = {Peng, Sainan and Chen, Yudong and Li, Yang and Fan, Guangyu and Xie, Xinhua and He, Feng and Tao, Zhensheng},
  journal = {Phys. Rev. Lett.},
  volume = {134},
  issue = {12},
  pages = {123203},
  numpages = {6},
  year = {2025},
  month = {Mar},
  publisher = {American Physical Society},
  doi = {10.1103/PhysRevLett.134.123203},
  url = {https://link.aps.org/doi/10.1103/PhysRevLett.134.123203}
}

@article{PhysRevLett.117.203001,
  title = {Role of Excited States In High-order Harmonic Generation},
  author = {Beaulieu, S. and Camp, S. and Descamps, D. and Comby, A. and Wanie, V. and Petit, S. and L\'egar\'e, F. and Schafer, K. J. and Gaarde, M. B. and Catoire, F. and Mairesse, Y.},
  journal = {Phys. Rev. Lett.},
  volume = {117},
  issue = {20},
  pages = {203001},
  numpages = {5},
  year = {2016},
  month = {Nov},
  publisher = {American Physical Society},
  doi = {10.1103/PhysRevLett.117.203001},
  url = {https://link.aps.org/doi/10.1103/PhysRevLett.117.203001}
}

@article{PhysRevA.96.023418,
  title = {Numerical observation of two sets of low-order harmonics near the ionization threshold},
  author = {Xiong, Wei-Hao and Jin, Jian-Zhao and Peng, Liang-You and Gong, Qihuang},
  journal = {Phys. Rev. A},
  volume = {96},
  issue = {2},
  pages = {023418},
  numpages = {6},
  year = {2017},
  month = {Aug},
  publisher = {American Physical Society},
  doi = {10.1103/PhysRevA.96.023418},
  url = {https://link.aps.org/doi/10.1103/PhysRevA.96.023418}
}

@article{Camp_2018,
doi = {10.1088/1361-6455/aaac12},
url = {https://doi.org/10.1088/1361-6455/aaac12},
year = {2018},
month = {feb},
publisher = {IOP Publishing},
volume = {51},
number = {6},
pages = {064001},
author = {Camp, Seth and Beaulieu, Samuel and Schafer, Kenneth J and Gaarde, Mette B},
title = {Resonantly-initiated quantum trajectories and their role in the generation of near-threshold harmonics},
journal = {Journal of Physics B: Atomic, Molecular and Optical Physics}
}

@incollection{kulander1993dynamics,
  title={Dynamics of short-pulse excitation, ionization and harmonic conversion},
  author={Kulander, KC and Schafer, KJ and Krause, JL},
  booktitle={Super-intense laser-atom physics},
  pages={95--110},
  year={1993},
  url={https://doi.org/10.1007/978-1-4615-7963-2_1},
  publisher={Springer}
}

@article{eichmann2009acceleration,
  title={Acceleration of neutral atoms in strong short-pulse laser fields},
  author={Eichmann, Ulli and Nubbemeyer, T and Rottke, H and Sandner, W},
  journal={Nature},
  volume={461},
  number={7268},
  pages={1261--1264},
  year={2009},
  publisher={Nature Publishing Group UK London}
}

@article{PhysRevLett.101.233001,
  title = {Strong-Field Tunneling without Ionization},
  author = {Nubbemeyer, T. and Gorling, K. and Saenz, A. and Eichmann, U. and Sandner, W.},
  journal = {Phys. Rev. Lett.},
  volume = {101},
  issue = {23},
  pages = {233001},
  numpages = {4},
  year = {2008},
  month = {Dec},
  publisher = {American Physical Society},
  doi = {10.1103/PhysRevLett.101.233001},
  url = {https://link.aps.org/doi/10.1103/PhysRevLett.101.233001}
}

@article{PhysRevLett.110.203002,
  title = {Observing Rydberg Atoms to Survive Intense Laser Fields},
  author = {Eichmann, U. and Saenz, A. and Eilzer, S. and Nubbemeyer, T. and Sandner, W.},
  journal = {Phys. Rev. Lett.},
  volume = {110},
  issue = {20},
  pages = {203002},
  numpages = {5},
  year = {2013},
  month = {May},
  publisher = {American Physical Society},
  doi = {10.1103/PhysRevLett.110.203002},
  url = {https://link.aps.org/doi/10.1103/PhysRevLett.110.203002}
}

@article{PhysRevLett.118.013003,
  title = {Unified Time and Frequency Picture of Ultrafast Atomic Excitation in Strong Laser Fields},
  author = {Zimmermann, H. and Patchkovskii, S. and Ivanov, M. and Eichmann, U.},
  journal = {Phys. Rev. Lett.},
  volume = {118},
  issue = {1},
  pages = {013003},
  numpages = {5},
  year = {2017},
  month = {Jan},
  publisher = {American Physical Society},
  doi = {10.1103/PhysRevLett.118.013003},
  url = {https://link.aps.org/doi/10.1103/PhysRevLett.118.013003}
}

@article{PhysRevA.107.043106,
  title = {Correlation analysis of frustrated tunneling ionization},
  author = {Ivanov, I. A. and Kheifets, A. S. and Kim, Kyung Taec},
  journal = {Phys. Rev. A},
  volume = {107},
  issue = {4},
  pages = {043106},
  numpages = {9},
  year = {2023},
  month = {Apr},
  publisher = {American Physical Society},
  doi = {10.1103/PhysRevA.107.043106},
  url = {https://link.aps.org/doi/10.1103/PhysRevA.107.043106}
}

@article{Zhao:19,
author = {Yong Zhao and Yueming Zhou and Jintai Liang and Zhexuan Zeng and Qinghua Ke and Yali Liu and Min Li and Peixiang Lu},
journal = {Opt. Express},
number = {15},
pages = {21689--21700},
publisher = {Optica Publishing Group},
title = {Frustrated tunneling ionization in the elliptically polarized strong laser fields},
volume = {27},
month = {Jul},
year = {2019},
url = {https://opg.optica.org/oe/abstract.cfm?URI=oe-27-15-21689},
doi = {10.1364/OE.27.021689},
}

@article{Liu:21,
author = {Mingqing Liu and Songpo Xu and Shilin Hu and Wilhelm Becker and Wei Quan and Xiaojun Liu and Jing Chen},
journal = {Optica},
number = {6},
pages = {765--770},
publisher = {Optica Publishing Group},
title = {Electron dynamics in laser-driven atoms near the continuum threshold},
volume = {8},
month = {Jun},
year = {2021},
url = {https://opg.optica.org/optica/abstract.cfm?URI=optica-8-6-765},
doi = {10.1364/OPTICA.418636},
}

@article{Landsman_2013,
doi = {10.1088/1367-2630/15/1/013001},
url = {https://doi.org/10.1088/1367-2630/15/1/013001},
year = {2013},
month = {jan},
publisher = {IOP Publishing},
volume = {15},
number = {1},
pages = {013001},
author = {Landsman, A S and Pfeiffer, A N and Hofmann, C and Smolarski, M and Cirelli, C and Keller, U},
title = {Rydberg state creation by tunnel ionization},
journal = {New Journal of Physics}
}

@article{PhysRevA.51.R1750,
  title = {Time profile of harmonics generated by a single atom in a strong electromagnetic field},
  author = {Antoine, Philippe and Piraux, Bernard and Maquet, Alfred},
  journal = {Phys. Rev. A},
  volume = {51},
  issue = {3},
  pages = {R1750--R1753},
  numpages = {0},
  year = {1995},
  month = {Mar},
  publisher = {American Physical Society},
  doi = {10.1103/PhysRevA.51.R1750},
  url = {https://link.aps.org/doi/10.1103/PhysRevA.51.R1750}
}

@article{pp22-qc27,
  title = {High-order harmonic generation in argon driven by short laser pulses: Effects of post-pulse propagation and windowing},
  author = {Bondy, Aaron T. and Bartschat, Klaus},
  journal = {Phys. Rev. A},
  volume = {113},
  issue = {6},
  pages = {063108},
  numpages = {9},
  year = {2026},
  month = {Jun},
  publisher = {American Physical Society},
  doi = {10.1103/pp22-qc27},
  url = {https://link.aps.org/doi/10.1103/pp22-qc27}
}

@article{He:26,
author = {Yu He and Eleni Tzima and Gergana D. Borisova and Lina Hedewig and Noah Wach and Harijyoti Mandal and Christian Ott and Thomas Pfeifer},
journal = {Opt. Lett.},
number = {7},
pages = {1752--1755},
publisher = {Optica Publishing Group},
title = {Laser-dressed ionic states in high-harmonic generation in helium},
volume = {51},
month = {Apr},
year = {2026},
url = {https://opg.optica.org/ol/abstract.cfm?URI=ol-51-7-1752},
doi = {10.1364/OL.590912},
}

@article{Yue:26,
author = {Zhongyao Yue and Yu He and Hongtao Hu and Chunyuan Bai and Chun Huang and Mette B. Gaarde and Candong Liu and Ruxin Li and Peng Peng},
journal = {Photon. Res.},
number = {6},
pages = {2428--2437},
publisher = {Optica Publishing Group},
title = {Mapping coupling pathways in attosecond XUV transient four-wave mixing of helium},
volume = {14},
month = {Jun},
year = {2026},
url = {https://opg.optica.org/prj/abstract.cfm?URI=prj-14-6-2428},
doi = {10.1364/PRJ.588447},
}

@article{Suzuki:06,
author = {Masayuki Suzuki and Motoyoshi Baba and Rashid Ganeev and Hiroto Kuroda and Tsuneyuki Ozaki},
journal = {Opt. Lett.},
number = {22},
pages = {3306--3308},
publisher = {Optica Publishing Group},
title = {Anomalous enhancement of a single high-order harmonic by using a laser-ablation tin plume at 47 nm},
volume = {31},
month = {Nov},
year = {2006},
url = {https://opg.optica.org/ol/abstract.cfm?URI=ol-31-22-3306},
doi = {10.1364/OL.31.003306},
}

@article{Ganeev:06,
author = {Rashid A. Ganeev and Masayuki Suzuki and Motoyoshi Baba and Hiroto Kuroda and Tsuneyuki Ozaki},
journal = {Opt. Lett.},
number = {11},
pages = {1699--1701},
publisher = {Optica Publishing Group},
title = {Strong resonance enhancement of a single harmonic generated in the extreme ultraviolet range},
volume = {31},
month = {Jun},
year = {2006},
url = {https://opg.optica.org/ol/abstract.cfm?URI=ol-31-11-1699},
doi = {10.1364/OL.31.001699},
}

@article{PhysRevLett.129.173202,
  title = {Role of Spin-Orbit Coupling in High-Order Harmonic Generation Revealed by Supercycle Rydberg Trajectories},
  author = {Mayer, N. and Beaulieu, S. and Jim\'enez-Gal\'an, \'A. and Patchkovskii, S. and Kornilov, O. and Descamps, D. and Petit, S. and Smirnova, O. and Mairesse, Y. and Ivanov, M. Y.},
  journal = {Phys. Rev. Lett.},
  volume = {129},
  issue = {17},
  pages = {173202},
  numpages = {5},
  year = {2022},
  month = {Oct},
  publisher = {American Physical Society},
  doi = {10.1103/PhysRevLett.129.173202},
  url = {https://link.aps.org/doi/10.1103/PhysRevLett.129.173202}
}

@article{RevModPhys.87.637,
  title = {Optical atomic clocks},
  author = {Ludlow, Andrew D. and Boyd, Martin M. and Ye, Jun and Peik, E. and Schmidt, P. O.},
  journal = {Rev. Mod. Phys.},
  volume = {87},
  issue = {2},
  pages = {637--701},
  numpages = {65},
  year = {2015},
  month = {Jun},
  publisher = {American Physical Society},
  doi = {10.1103/RevModPhys.87.637},
  url = {https://link.aps.org/doi/10.1103/RevModPhys.87.637}
}

@article{PhysRevA.102.062809,
  title = {Photoelectron spectroscopy of laser-dressed atomic helium},
  author = {Meister, Severin and Bondy, Aaron and Schnorr, Kirsten and Augustin, Sven and Lindenblatt, Hannes and Trost, Florian and Xie, Xinhua and Braune, Markus and Treusch, Rolf and Manschwetus, Bastian and Schirmel, Nora and Redlin, Harald and Douguet, Nicolas and Pfeifer, Thomas and Bartschat, Klaus and Moshammer, Robert},
  journal = {Phys. Rev. A},
  volume = {102},
  issue = {6},
  pages = {062809},
  numpages = {6},
  year = {2020},
  month = {Dec},
  publisher = {American Physical Society},
  doi = {10.1103/PhysRevA.102.062809},
  url = {https://link.aps.org/doi/10.1103/PhysRevA.102.062809}
}

@article{PhysRevA.49.2117,
  title = {Theory of high-harmonic generation by low-frequency laser fields},
  author = {Lewenstein, M. and Balcou, Ph. and Ivanov, M. Yu. and L'Huillier, Anne and Corkum, P. B.},
  journal = {Phys. Rev. A},
  volume = {49},
  issue = {3},
  pages = {2117--2132},
  numpages = {0},
  year = {1994},
  month = {Mar},
  publisher = {American Physical Society},
  doi = {10.1103/PhysRevA.49.2117},
  url = {https://link.aps.org/doi/10.1103/PhysRevA.49.2117}
}

@article{PhysRevLett.64.1107,
  title = {Nonlinear optical processes using electromagnetically induced transparency},
  author = {Harris, S. E. and Field, J. E. and Imamo\ifmmode \breve{g}\else \u{g}\fi{}lu, A.},
  journal = {Phys. Rev. Lett.},
  volume = {64},
  issue = {10},
  pages = {1107--1110},
  numpages = {0},
  year = {1990},
  month = {Mar},
  publisher = {American Physical Society},
  doi = {10.1103/PhysRevLett.64.1107},
  url = {https://link.aps.org/doi/10.1103/PhysRevLett.64.1107}
}

@article{PhysRevLett.66.2593,
  title = {Observation of electromagnetically induced transparency},
  author = {Boller, K.-J. and Imamo\ifmmode \breve{g}\else \u{g}\fi{}lu, A. and Harris, S. E.},
  journal = {Phys. Rev. Lett.},
  volume = {66},
  issue = {20},
  pages = {2593--2596},
  numpages = {0},
  year = {1991},
  month = {May},
  publisher = {American Physical Society},
  doi = {10.1103/PhysRevLett.66.2593},
  url = {https://link.aps.org/doi/10.1103/PhysRevLett.66.2593}
}

\clearpage
\includepdf[
    pages=1,
    landscape=false,
    turn=false
]{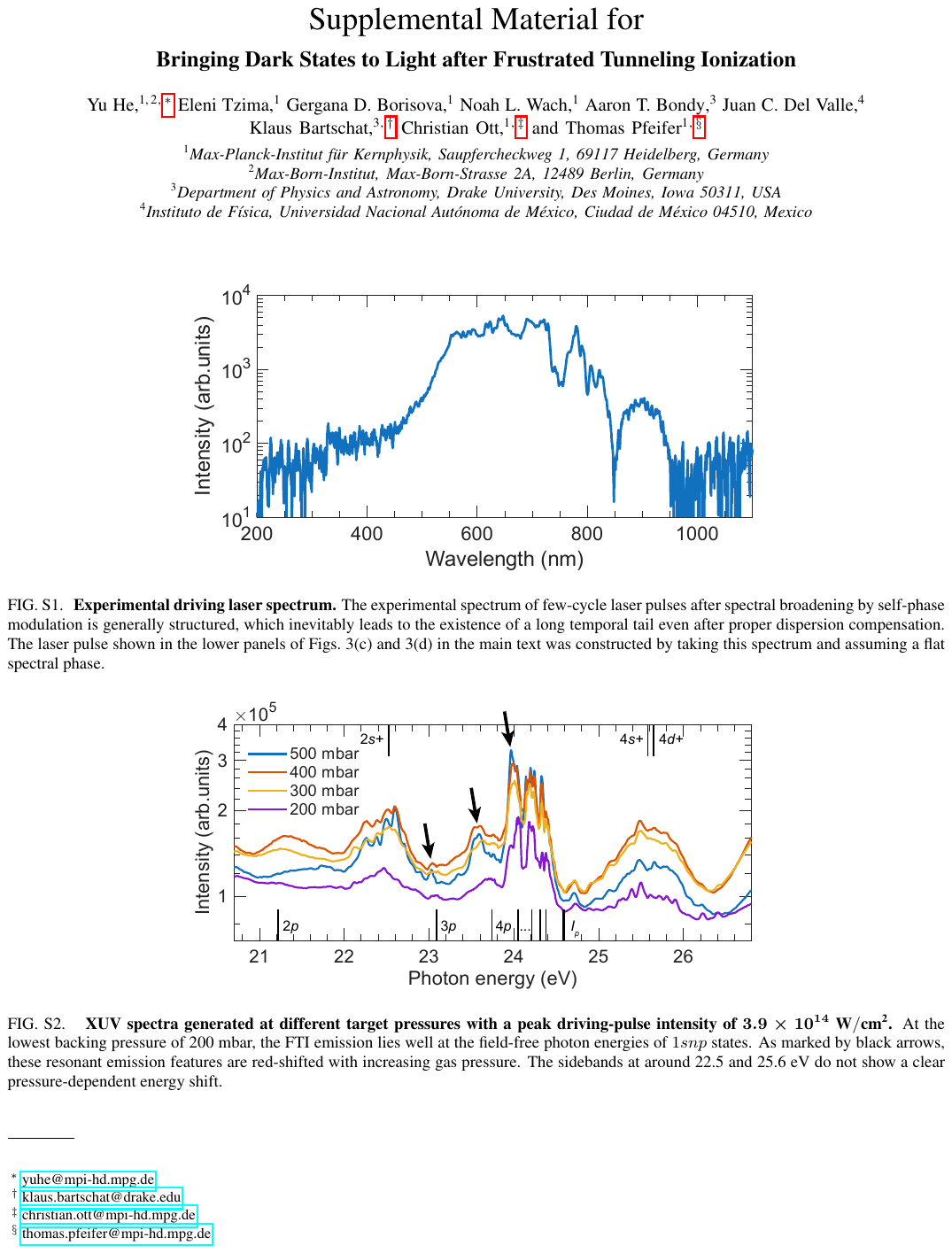}

\clearpage
\includepdf[
    pages=2,
    landscape=false,
    turn=false
]{HeDarkStatestoLight_SM.pdf}

\end{document}